\documentclass[conference, 9pt]{IEEEtran}
\IEEEoverridecommandlockouts
\usepackage{cite}
\usepackage{amsmath,amssymb,amsfonts}
\usepackage{algorithmic}
\usepackage{graphicx}
\usepackage{textcomp}
\usepackage{array}
\usepackage[table,xcdraw]{xcolor}
\usepackage{booktabs}

\usepackage{subfig}

\def\BibTeX{{\rm B\kern-.05em{\sc i\kern-.025em b}\kern-.08em
    T\kern-.1667em\lower.7ex\hbox{E}\kern-.125emX}}
\begin{document}

\newcommand{\ml}[1]{{\color{red}\bf [Meng: #1]}}

\newcommand{\method}{SpecASR}

\title{SpecASR: Accelerating LLM-based Automatic Speech Recognition via Speculative Decoding}

\author{\IEEEauthorblockN{Linye Wei$^{1,2}$, Shuzhang Zhong$^{1,2}$, Songqiang Xu$^{1,3}$, Runsheng Wang$^{2,4,5}$, Ru Huang$^{2,4,5}$, Meng Li$^{1,2,5*}$}
\IEEEauthorblockA{$^1$Institute for Artificial Intelligence, Peking University, Beijing, China}
\IEEEauthorblockA{$^2$School of Integrated Circuits, Peking University, Beijing, China}
\IEEEauthorblockA{$^3$School of Software and Microelectronics, Peking University, Beijing, China}
\IEEEauthorblockA{$^4$Institute of Electronic Design Automation, Peking University, Wuxi, China}
\IEEEauthorblockA{$^5$Beijing Advanced Innovation Center for Integrated Circuits, Beijing, China}
\IEEEauthorblockA{meng.li@pku.edu.cn}
\thanks{This work was supported in part by NSFC under Grant 62495102 and Grant 92464104, in part by the National Key Research and Development Program under Grant 2024YFB4505004, in part by Beijing Municipal Science and Technology Program under Grant Z241100004224015, and in part by 111 Project under Grant B18001.}}


\maketitle


\begin{abstract}
    Large language model (LLM)-based automatic speech recognition (ASR) has recently attracted a lot of attention due to its high recognition accuracy and enhanced multi-dialect support. However, the high decoding latency of LLMs challenges the real-time ASR requirements. Although speculative decoding has been explored for better decoding efficiency, they usually ignore the key characteristics of the ASR task and achieve limited speedup.  To further reduce the real-time ASR latency, in this paper, we propose a novel speculative decoding framework specialized for ASR, dubbed~\method. \method~is developed based on our core observation that ASR decoding is audio-conditioned, which results in high output alignment between small and large ASR models, even given output mismatches in intermediate decoding steps. Therefore, \method~features an adaptive draft sequence generation process that dynamically modifies the draft sequence length to maximize the token acceptance length. \method~further proposes a draft sequence recycling strategy that reuses the previously generated draft sequence to reduce the draft ASR model latency. Moreover, a two-pass sparse token tree generation algorithm is also proposed to balance the latency of draft and target ASR models. With extensive experimental results, we demonstrate \method~achieves 3.04×–3.79× and 1.25×–1.84× speedup over the baseline autoregressive decoding and speculative decoding, respectively, without any loss in recognition accuracy.

\end{abstract}


\section{Introduction}
\label{sec:intro}

Large language models (LLMs) like GPT~\cite{achiam2023gpt} and Llama~\cite{touvron2023llama} have demonstrated powerful capabilities across a wide range of applications such as text generation, question answering \cite{zhuang2023toolqa}, and embodied AI \cite{brohan2023rt}. Recently, LLM has also been applied to automatic speech recognition (ASR) \cite{wu2023decoder,seide2024speech,bai2024seed,fang2024llama,fathullah2024prompting,chen2024bestow}. By incorporating an audio encoder for speech embedding extraction, LLM-based ASR models have demonstrated impressive recognition accuracy and enhanced support for a broader range of languages, dialects, and accents \cite{bai2024seed}.

\begin{figure}[!tb]
    \centering    
    \subfloat[][]{
	\includegraphics[width=0.95\linewidth]{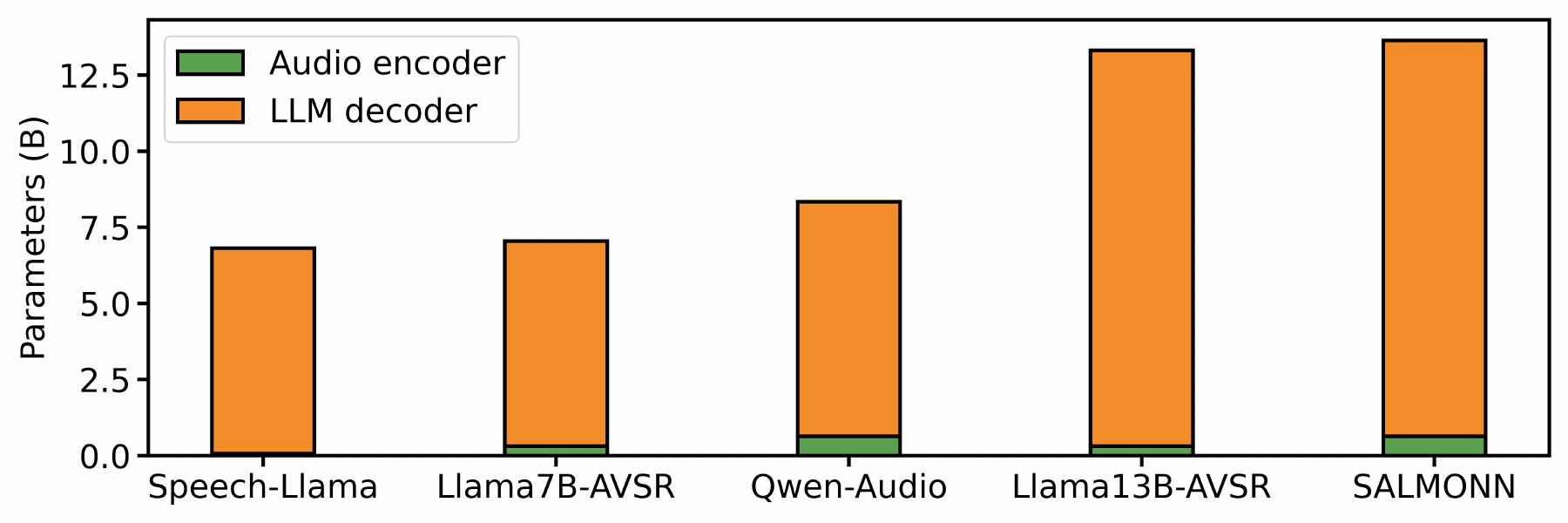}
        \label{fig:subfig1a}
    }
    \hspace{0.01\textwidth}
    \subfloat[][]{
	\includegraphics[width=0.95\linewidth]{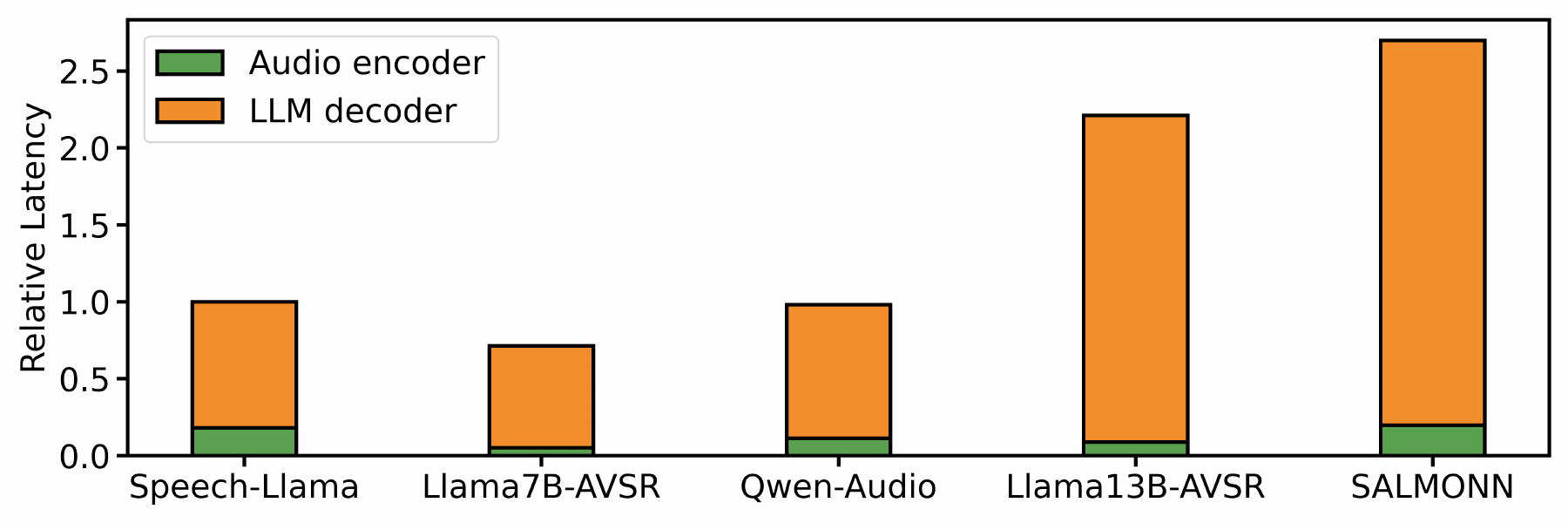}
        \label{fig:subfig1b}
    }
    \caption{(a) Parameter ratio and (b) relative latency of audio encoders and LLM decoders in LLM-based ASR models.}
    \label{fig1}
\end{figure}

However, the improvement of LLM-based ASR is achieved at the cost of much longer decoding latency as LLM significantly increases the model size and compute of the ASR models. For example, the LLMs in BESTOW \cite{chen2024bestow}, Speech-Llama \cite{fathullah2024prompting}, and Seed-ASR \cite{bai2024seed} have 1.1B, 7B, and $>$10B parameters, respectively. In contrast, the model size of an audio encoder is generally under 1B, and even below 100M. The autoregressive decoding nature of LLM further exacerbates the computation latency. As a result, as shown in Fig.~\ref{fig1}, the LLM decoder incurs much larger inference latency compared to the audio encoder, becoming the major efficiency bottleneck of ASR.



To improve the LLM decoding efficiency, speculative decoding has been extensively studied in the context of natural language processing tasks \cite{chen2023accelerating,leviathan2023fast,miao2023specinfer,cai2024medusa,li2024eagle,xiong2024dyspec} and has also recently been introduced to ASR models \cite{segal2024whisper}. The core idea of speculative decoding is first to employ a small ASR model to generate draft output tokens autoregressively. Then, the target large ASR model is used to verify the draft sequences altogether. The draft tokens that match the outputs of the target ASR model are accepted, which improves the efficiency of the target ASR model without sacrificing recognition quality. Though effective, existing works often directly apply speculative decoding to ASR but ignore its key characteristics \cite{segal2024whisper,gandhi2023distil}. Hence, they only achieve limited speedup, e.g., $\sim 1.5\times$, which may still be insufficient to meet the tight real-time constraints. 

In this paper, we observe that ASR is an audio-conditioned generation task, which results in two key characteristics: on one hand, unlike many text tasks, the outputs of large and small ASR models are often highly aligned; on the other hand, even given output mismatches between the large and small models, the downstream decoding of two models can still be aligned due to the audio input. Based on these observations, we propose a highly efficient speculative decoding framework, dubbed~\method, specialized for LLM-based ASR. \method~features 3 key techniques for efficient ASR, including an adaptive single-sequence prediction algorithm to maximize the decoding-acceptance ratio in each verification step, a draft sequence recycling strategy to reuse the unaccepted draft sequence and reduce draft generation latency, and a two-pass sparse token tree algorithm to balance the latency of draft generation and target verification. Our main contributions can be summarized as follows:
\begin{itemize}
    \item We observe that ASR decoding is audio conditioned and propose SpecASR to improve the ASR decoding efficiency.
    \item We propose 3 novel speculative decoding techniques, including adaptive single-sequence prediction, draft sequence recycling, and two-pass sparse token tree to leverage ASR characteristics.
    \item With extensive experimental results, SpecASR achieves 3.04×–3.79× and 1.25×–1.84× latency reduction compared to the baseline autoregressive and speculative decoding algorithms, respectively, with iso-accuracy.
\end{itemize}



\section{Background}

\subsection{LLM-based ASR Models}
\begin{figure}[!tb]
    \centering
    \includegraphics[width=0.95\linewidth]{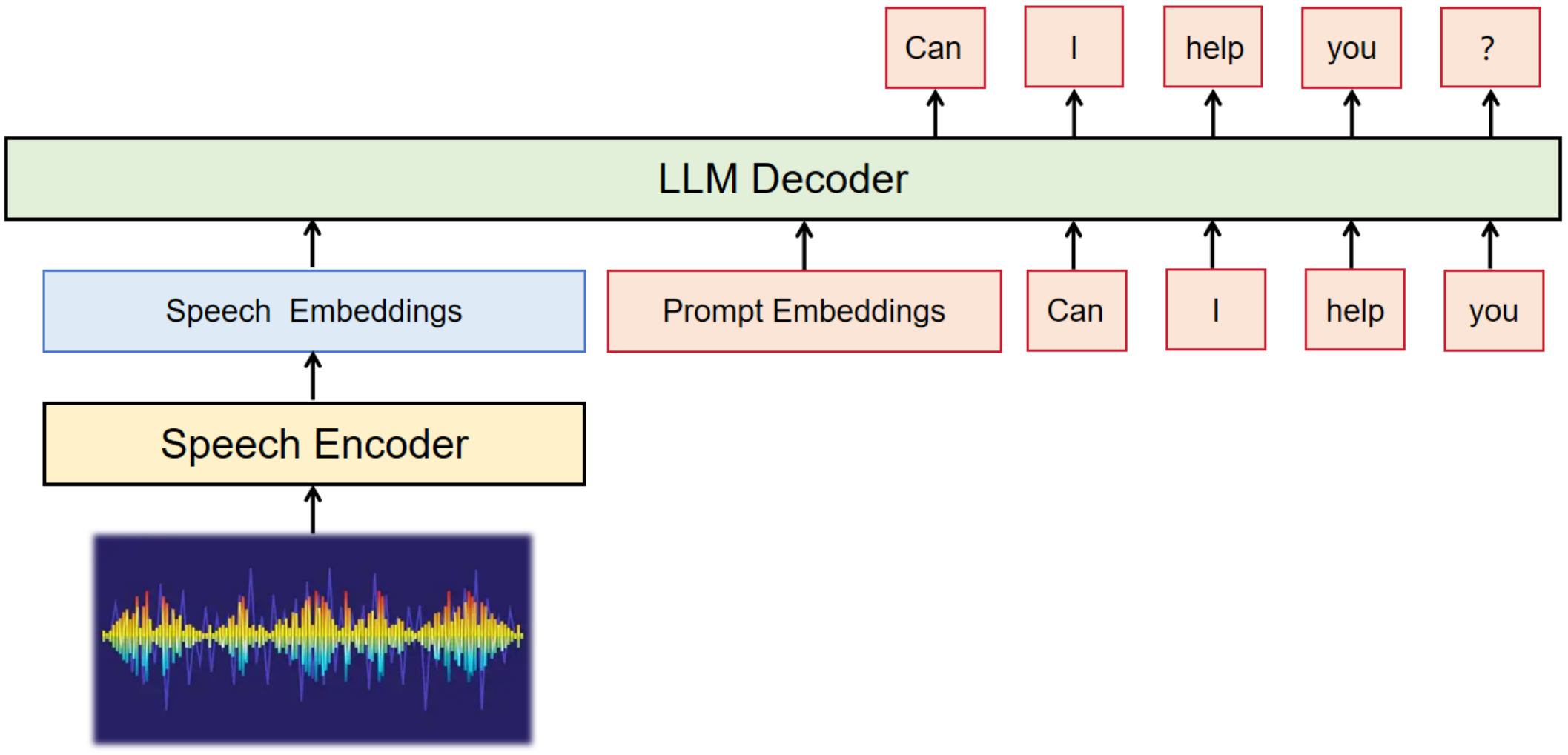}
    \caption{Cascaded architecture of LLM-based ASR models.}
    \label{fig2}
\end{figure}

The integration of LLMs into ASR tasks has demonstrated significant advantages in recognition accuracy and contextual awareness. LLM-based ASR models utilize a cascaded architecture consisting of an audio encoder and an LLM decoder, as illustrated in Fig.~\ref{fig2}. The audio encoder is responsible for feature extraction and dimensional transformation of the speech input, while the LLM decoder generates text transcriptions through autoregressive decoding.

\paragraph{Audio Encoder} In LLM-based ASR models, audio encoders serve as bridges connecting audio signals to LLMs originally designed for text tasks. The processing is conducted in two stages. The first stage encompasses feature extraction and compression of speech frames by the audio encoder as the number of speech frames significantly exceeds the text tokens needed to convey the same semantics. In the following stage, the extracted features are subjected to stacking, enabling projection into the hidden dimensions of LLMs and allowing for prefilling alongside text prompts. Recent studies leverage established ASR architectures such as Conformer\cite{gulati2020conformer} or Whisper\cite{radford2023robust} as audio encoders. As illustrated in Fig.~\ref{fig1}, audio encoders typically account for a small fraction of the model.

\paragraph{LLM Decoder} The LLM decoder processes audial embeddings alongside text prompts to perform autoregressive decoding of speech transcriptions. Prior studies have shown that text LLMs, such as GPT\cite{achiam2023gpt}, Llama\cite{touvron2023llama} and Qwen\cite{bai2023qwen}, can be effectively utilized in ASR tasks by employing either pre-trained models directly or fine-tuning them with a limited number of parameters via LoRA\cite{hu2021lora}. 
However, the complex architecture of LLMs and the autoregressive decoding process result in a long latency, making the decoder a major bottleneck and hence, the main focus of our work.

\subsection{Speculative Decoding}
\begin{figure}[!tb]
    \centering
    \includegraphics[width=0.95\linewidth]{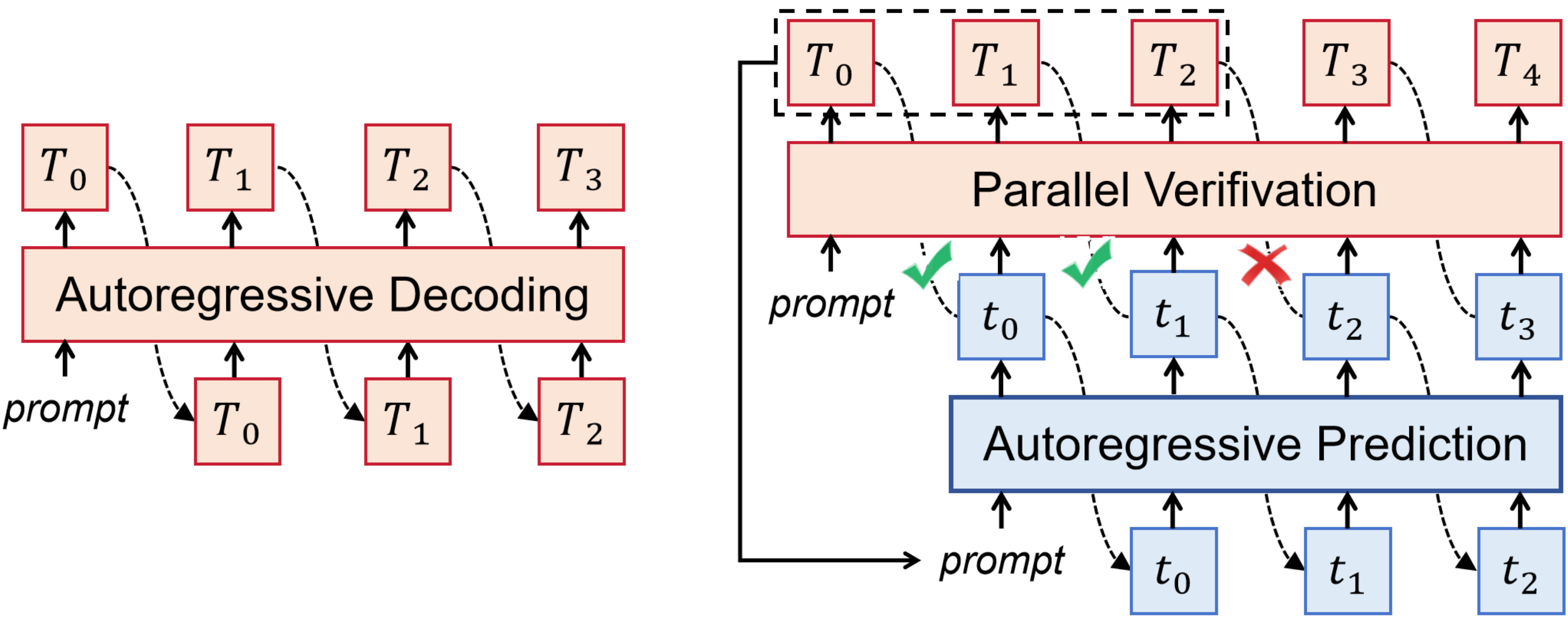}
    \caption{Autoregressive (left) and speculative (right) decoding.}
    \label{fig3}
\end{figure}

Speculative decoding \cite{miao2023specinfer,chen2023accelerating,leviathan2023fast,cai2024medusa,zhong2024propd,li2024eagle2,chen2024sequoia,yang2024multi,li2024eagle} incorporates lightweight draft models to enhance the decoding speed of target models through the ``Draft-then-Verify'' paradigm. As shown in Fig.~\ref{fig3}, the draft model first efficiently generates predictions for multiple forthcoming tokens autoregressively. Then, the target model concurrently verifies all tokens within the draft sequence, terminating at the first token that fails verification and reverting to the draft model for the next round of predictions. Hence, speculative decoding guarantees lossless acceleration of target models' decoding process.

\begin{table*}[!tb]
\caption{Comparison between our methods with representative speculative decoding methods.}
\label{tab1}
\centering
\setlength{\tabcolsep}{8pt} 
\renewcommand{\arraystretch}{1.2}
\resizebox{\linewidth}{!}{
\begin{tabular}{c|ccccc}
\toprule
Method & Draft Generation Efficiency & Target Verification Efficiency & Draft Sequence Length & Target Accept Rate & Flexibility \\
\midrule
Single Sequence \cite{chen2023accelerating,leviathan2023fast} & high & low & medium & low & medium \\
\hline
Fixed Tree \cite{miao2023specinfer,li2024eagle,yang2024multi} & low & high & low & medium & low \\
\hline
Dynamic Tree \cite{cai2024medusa,zhong2024propd,li2024eagle2,chen2024sequoia} & low & high & low & high & high \\
\hline
\rowcolor{gray!20}
Ours & high & high & high & high & high \\
\bottomrule
\end{tabular}
}
\end{table*}

\begin{figure}[!tb]
    \centering
    \includegraphics[width=0.7\linewidth]{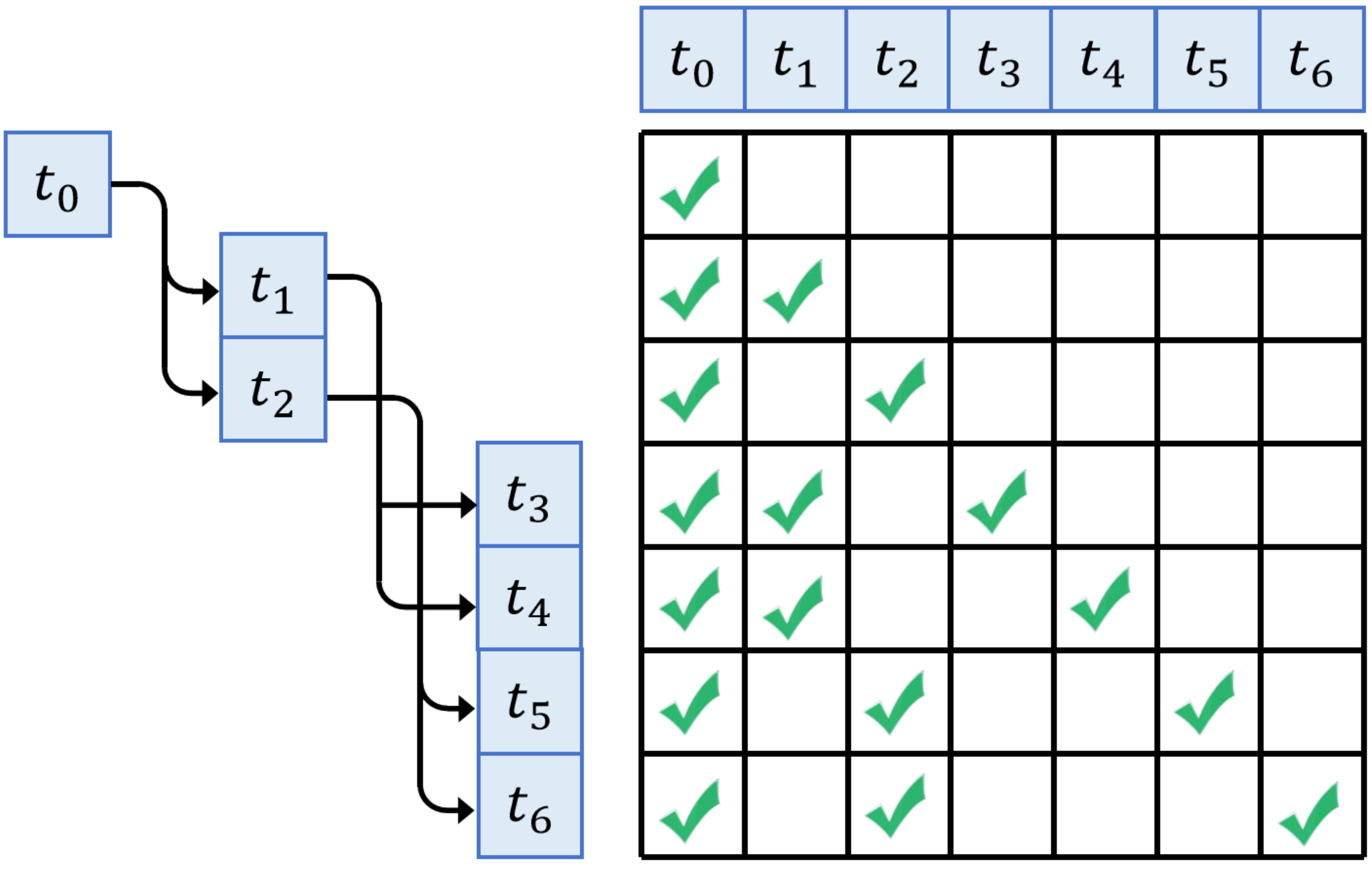}
    \caption{2D attention mask for draft token tree.}
    \label{fig13}
\end{figure}

Tree-structure speculative decoding like SpecInfer\cite{miao2023specinfer} creatively extends a single draft sequence into a fixed-shape token tree that encompasses multiple candidate draft sequences. As shown in Fig.~\ref{fig13}, this method maintains multiple candidate tokens at each decoding step of the draft model, arranging them into a tree-like structure. During verification, the candidate sequences are expanded and merged to facilitate parallel processing. Leveraging a 2D attention mask, the target model effectively retrieves independent verification outcomes for each branch. By promoting greater diversity in predictions at each decoding step, tree-structure speculative decoding achieves a higher success rate during the target model's verification phase, thus accelerating the speculative process. Recent works, such as Medusa\cite{cai2024medusa} and ProPD\cite{zhong2024propd}, have further refined the construction of draft token trees by selecting token branches based on the probability distribution. This dynamic tree-structure speculative decoding increases the verification acceptance rate and improves flexibility when applied to various models and tasks.

However, the abandonment of draft sequences that fail verification in speculative decoding results in significant wastage of draft predictions. Furthermore, existing tree-structure speculative approaches primarily focus on enhancing the diversity of draft model decoding, limiting tree depths to prevent the prediction overhead from growing exponentially with the number of forward passes. In the context of ASR tasks, the high alignment between the decoding outputs of the draft model and the target model suggests that draft tokens have the potential for multiple reuses. Also, a token tree that emphasizes both constrained width expansion and proactive depth expansion may yield improved performance. Tab.~\ref{tab1} presents a comparison between our method and existing speculative decoding approaches.


\section{Motivation}
\label{sec:motivation}

In this section, we discuss the feasibility of applying speculative decoding to ASR tasks, examining the unique characteristics of the speculative decoding process within ASR models and the potential for efficiency improvements. Based on these analyses, we present three key observations that will motivate our subsequent work.

\vspace{5pt}

\begin{figure}[!tb]
    \centering    
    \subfloat[][]{
	\includegraphics[width=0.48\linewidth]{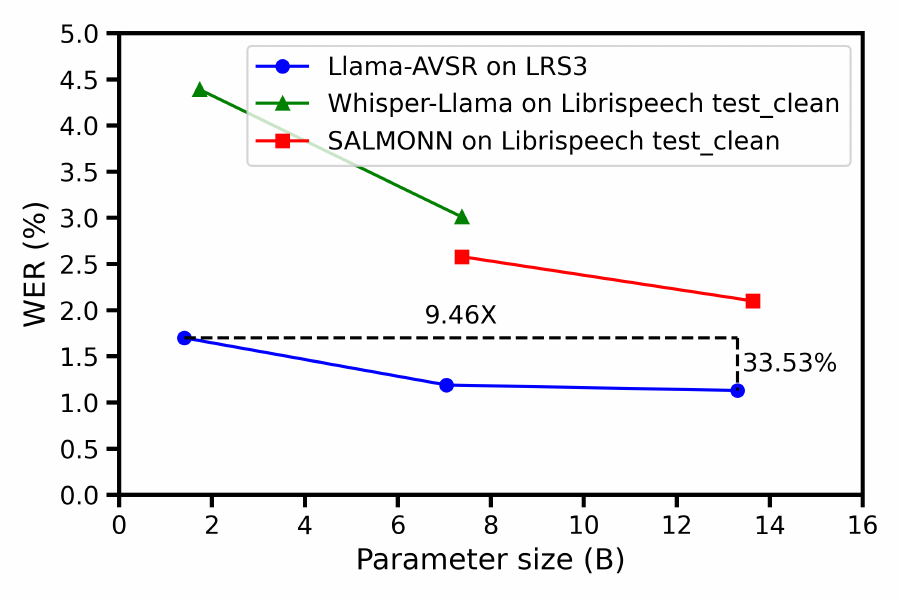}
        \label{fig:subfig4a}
    }
    \subfloat[][]{
	\includegraphics[width=0.48\linewidth]{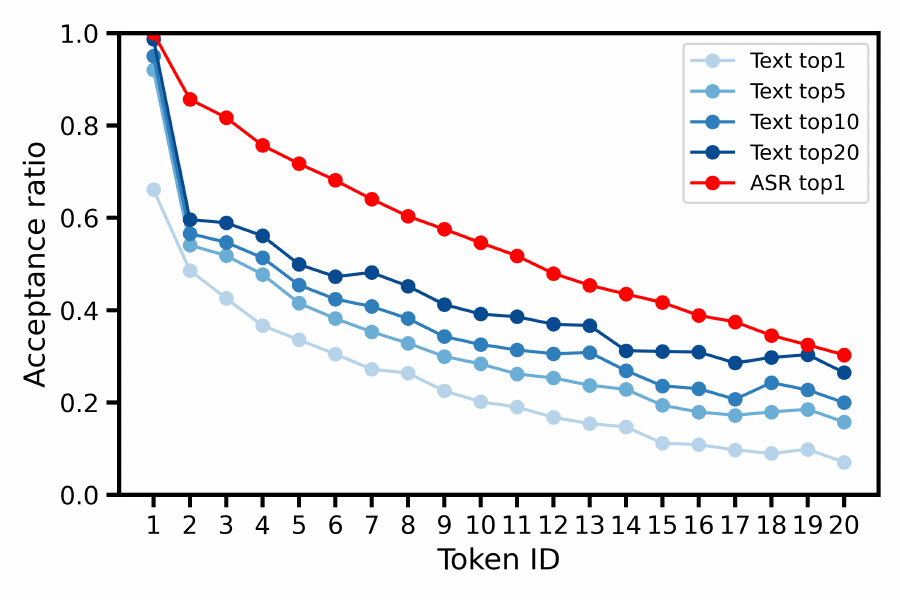}
        \label{fig:subfig4b}
    }
    \caption{(a) WERs of ASR models with multiple scales, (b) comparison of speculative acceptance with top-k logits in ASR/text tasks.}
    \label{fig4}
\end{figure}

\textbf{Observation 1: Applying speculative decoding to ASR tasks is effective and is expected to yield higher speedups.} As illustrated in Fig.~\ref{fig:subfig4a}, we evaluate the performance of ASR models with varying parameter sizes across multiple datasets\cite{panayotov2015librispeech,afouras2018lrs3}. Using word error rate (WER) as the primary evaluation metric, larger models exhibit a 20\%-33\% reduction in WER compared to smaller models, albeit with a significant increase in model size. While the performance of smaller models falls short of meeting user expectations in practical ASR applications, they show considerable potential within the context of speculative decoding, achieving WERs as low as 10\% or less.

However, speculative decoding for ASR tasks remains an area of limited exploration. Existing approaches typically apply single-sequence prediction or tree-structured speculative decoding techniques, originally developed for text tasks, to ASR models, resulting in modest speedups of only 1.5×–2×. As shown in Fig.~\ref{fig:subfig4b}, our findings demonstrate that draft models exhibit significantly improved performance in ASR tasks compared to text-based applications. By capitalizing on the strong consistency between the draft and target models, it is possible to employ longer prediction lengths and constrained token tree expansion, thereby improving the verification efficiency of the target model.

\vspace{5pt}

\begin{figure}[!tb]
    \centering    
    \subfloat[][]{
	\includegraphics[width=0.58\linewidth]{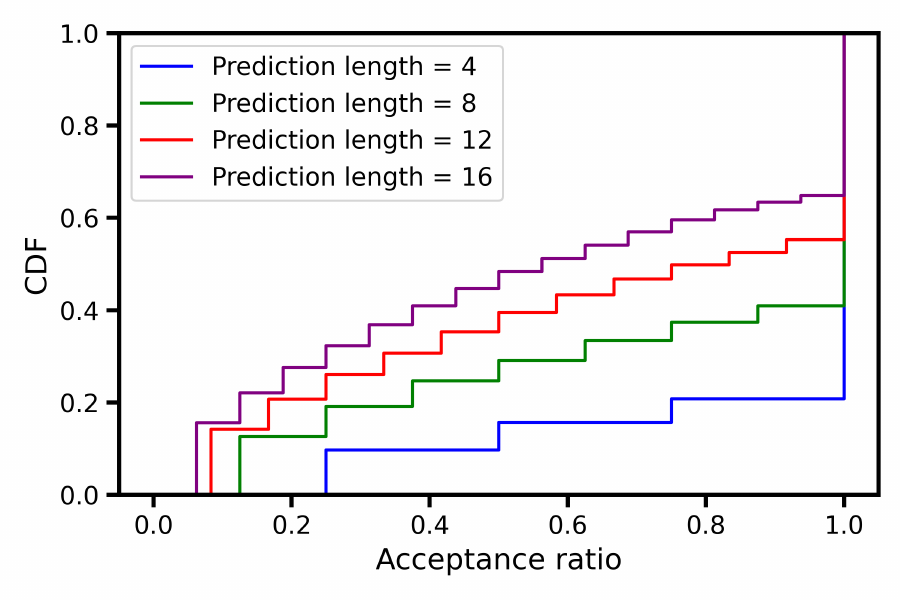}
        \label{fig:subfig5a}
    }
    \subfloat[][]{
	\includegraphics[width=0.36\linewidth]{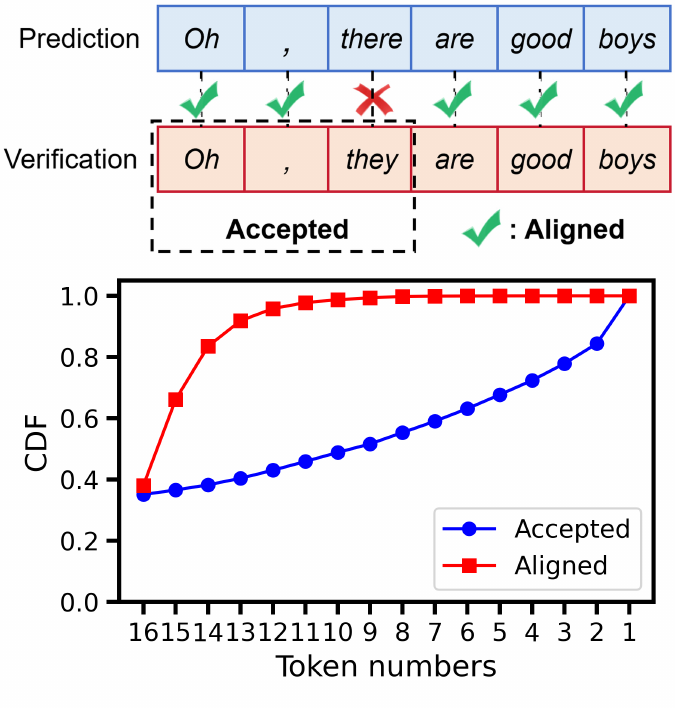}
        \label{fig:subfig5b}
    }
    \caption{(a) Acceptance ratio distribution with different prediction lengths, (b) high alignment of draft and target models in ASR tasks.}
    \label{fig5}
\end{figure}

\textbf{Observation 2: In ASR tasks, the acceptance rate of the draft model in speculative decoding exhibits substantial variation, with some draft sequences showing low acceptance rates yet strong alignment with verification sequences.} We analyze the distribution of acceptance rates under varying prediction lengths in speculative decoding, as shown in Fig.~\ref{fig:subfig5a}. A substantial proportion of predictions are fully accepted, highlighting the effectiveness of long-sequence draft decoding. However, the remaining acceptance rates are primarily concentrated at lower values, which can be attributed to variations in pronunciation and acoustic quality across specific speech segments. These factors lead to a localized error distribution, contributing to inefficiencies in draft predictions.

To address this issue, we are motivated to employ dynamic prediction lengths. Our observations reveal that, due to the role of the audio encoder in ASR tasks, which ensures the basic semantic coherence of the transcription process, ASR decoding is audio-conditioned. As shown in Fig.~\ref{fig:subfig5b}, unaccepted draft sequences demonstrate a high degree of alignment with the verification sequence. This suggests that the computational overhead of draft models can be significantly reduced by reusing unaccepted draft sequences.

\vspace{5pt}

\textbf{Observation 3: In speculative decoding, the relative contributions of the draft and target models to overall latency vary significantly across configurations, such as prediction length and model size.} Current tree-structured speculative approaches are based on the assumption that the verification process of the target model is the primary bottleneck in achieving decoding speedups with constrained prediction lengths. As a result, promoting diversity in the draft model's predictions to enhance verification acceptance rates has been shown to effectively accelerate decoding. However, ASR decoding benefits from longer prediction lengths to improve overall efficiency, highlighting the need for a thorough analysis of the computational load distribution between the draft and target models.

\begin{figure}[!tb]
\centerline{\includegraphics[width=0.48\textwidth]{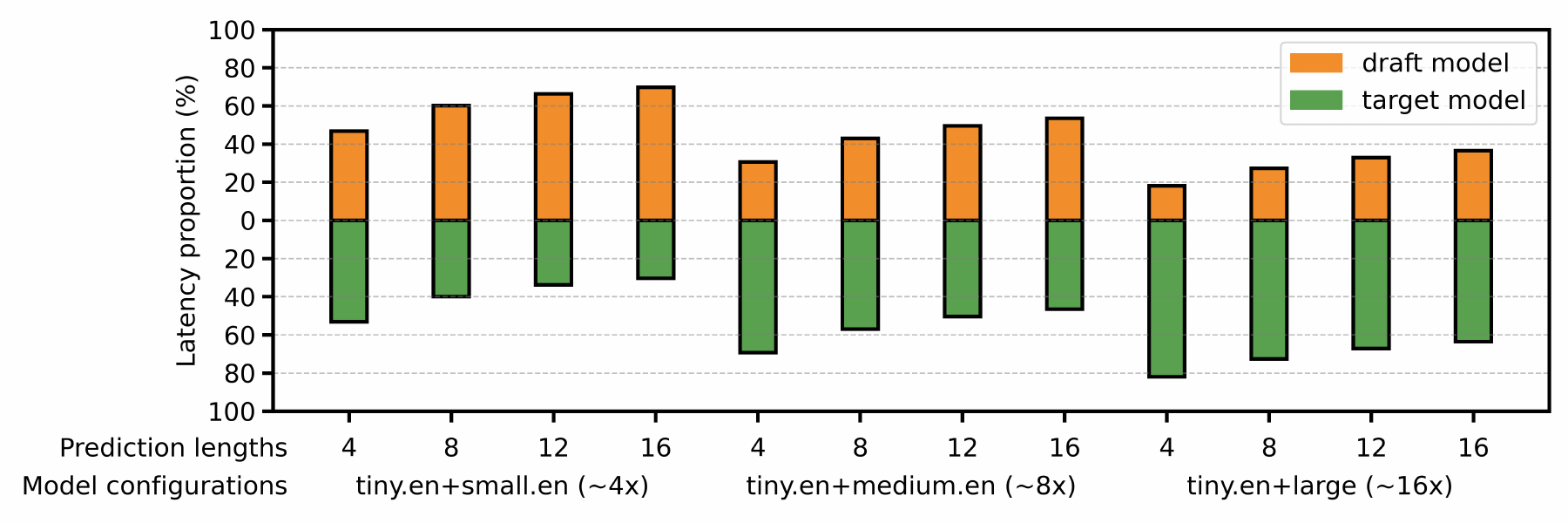}}
\caption{Decoding latency proportion on LibriSpeech clean-test.}
\label{fig6}
\end{figure}

As shown in Fig.~\ref{fig6}, we evaluate the proportion of latency contributed by draft prediction and target verification across varying prediction lengths and model configurations. As the prediction length increases, the draft model progressively becomes the dominant source of latency. In contrast, for a fixed prediction length, a significant size disparity between the draft and target models often results in the target model being the primary one. This suggests that, depending on specific task requirements, latency bottlenecks may arise from either model. Our proposed approach must effectively balance the trade-off between the draft and target models to achieve optimal speedups.

\section{SpecASR Framework}

\begin{figure}[!tb]
    \centerline{\includegraphics[width=0.95\linewidth]{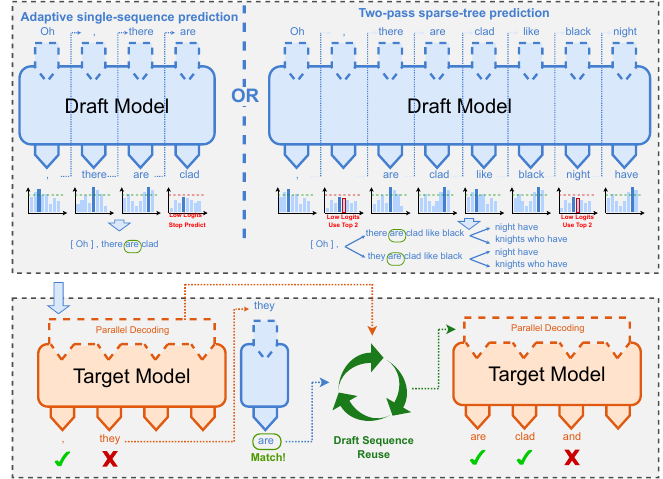}}
    \caption{The overview framework of SpecASR.}
    \label{fig7}
\end{figure}

Building on these analysis, we propose SpecASR, a novel framework designed to accelerate the decoding process of ASR models, as illustrated in Fig.~\ref{fig7}. SpecASR incorporates adaptive single-sequence prediction, dynamically adjusting the length of draft sequences in each prediction round to enhance verification efficiency. To mitigate the computational burden on the draft model, SpecASR incorporates a draft sequence recycling strategy. This approach leverages the opportunity to merge unaccepted tokens during decoding, dynamically adjusting the length of draft sequences in each regeneration phase. Furthermore, we introduce a two-pass sparse-tree prediction method as an alternative decoding strategy to accommodate diverse target model configurations in ASR tasks. This approach effectively balances latency between the draft and target models, enabling optimal speedup, particularly for larger-scale target models.

\subsection{Adaptive Single-sequence Prediction}

The strong alignment observed between the decoding processes of large and small models in ASR tasks motivates the adoption of longer prediction lengths to reduce the number of verification rounds. Unlike prior studies, which generally employ limited prediction lengths \cite{xiao2024parallelspec,liu2023online}, we extend the maximum prediction length to 24 tokens. This extension fully exploits the draft model's competitive decoding capabilities, enhancing overall efficiency. Before the draft sequence reaches the predefined length, we proactively identify positions with a high probability of verification failure during the prediction process. At these positions, the draft model's decoding is truncated, and the generated tokens are promptly passed to the target model for verification. An analysis of the logits distribution for accepted and rejected tokens during verification reveals a strong positive correlation between acceptance probability and normalized logits. Leveraging this insight, we introduce a threshold-based mechanism to truncate predictions with logits falling below a specified value, enabling early verification of the draft sequence. This adaptive approach to prediction length facilitates extended prediction and acceptance lengths while preserving the efficiency of the draft model.

\subsection{Draft Sequence Recycling Strategy}

\begin{figure}[!tb]
    \centerline{\includegraphics[width=0.95\linewidth]{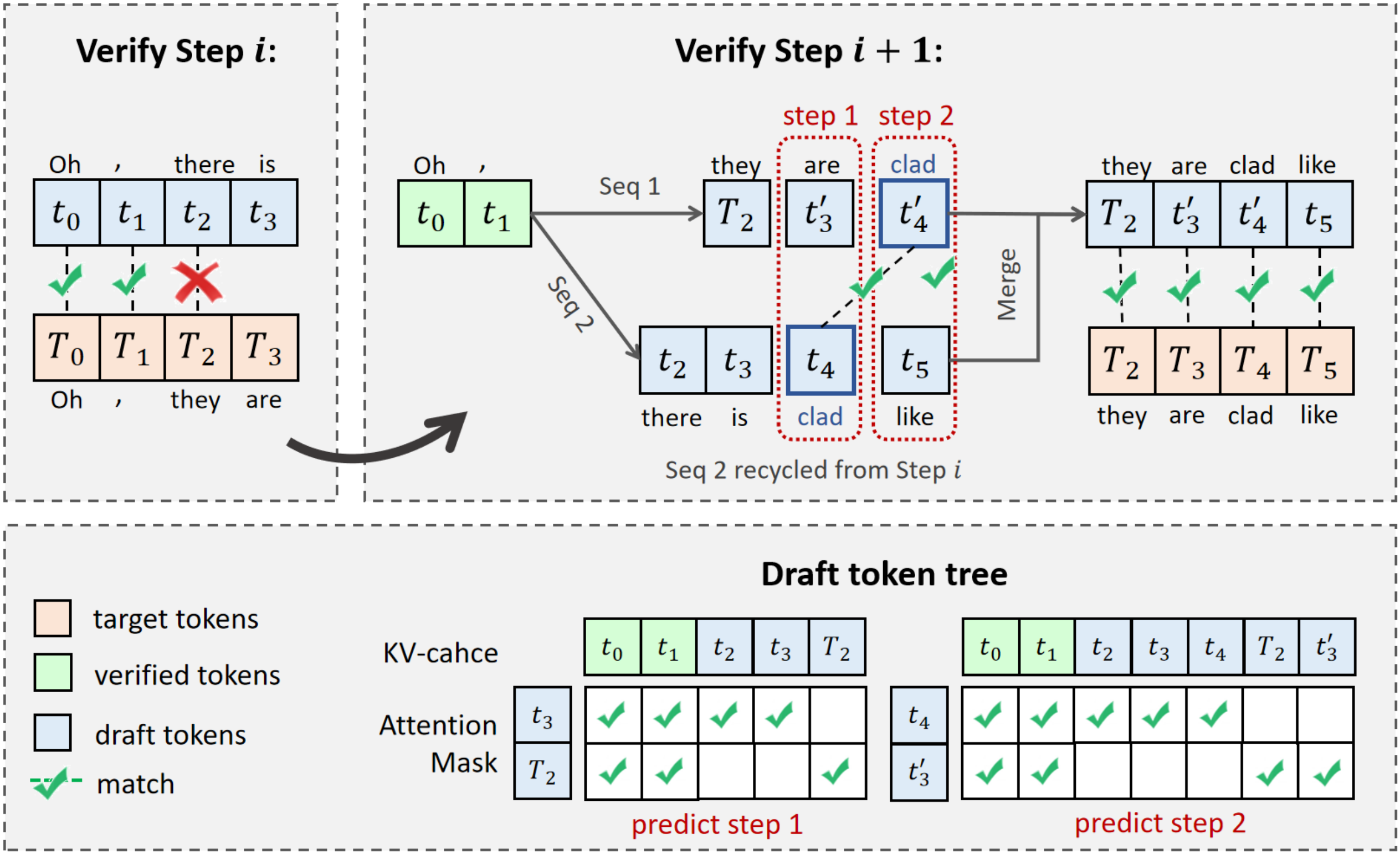}}
    \caption{Draft sequence recycling strategy.}
    \label{fig8}
\end{figure}

Given that the logits of unaccepted draft tokens span a wide range of values, even outputs with normalized logits exceeding the set threshold may fail verification, leading to the rejection of a substantial number of tokens. Building on the insights presented in Section III, we propose a strategy to recycle these draft tokens. As illustrated in Fig.~\ref{fig8}, unlike previous approaches that construct draft token trees to improve verification acceptance rates \cite{sun2024spectr,jeon2024recursive,cheng2024recurrent}, we employ a masked token tree to enable parallel prediction and token reuse within the draft model. Our recycling strategy is composed of two key stages: draft regeneration and merging.

In the first stage, the complete draft sequence submitted for verification is retained as sequence 1, while the partially verified tokens are stored as sequence 2 to construct a draft token tree. By applying an attention mask to the unaccepted tokens, the draft model performs parallel decoding, simultaneously extending the draft sequence and regenerating tokens that failed verification. This approach effectively conceals the regeneration delay within the ongoing predictions of the draft model. In the second stage, the positions of all tokens are tracked and marked. During each step of the regeneration process, the tokens generated in sequence 2 are compared with those in sequence 1 at corresponding or adjacent positions. Upon identifying a match, the two branches of the draft token tree are merged, enabling the reuse of previously decoded content and alleviating the prediction workload on the draft model.

\subsection{Two-pass Sparse-tree Prediction}

The preceding discussion builds on Observation 3 in Section III, which highlights that the draft model dominates the latency in speculative decoding with extended draft prediction lengths. However, this observation also notes that as the parameter size of the target model increases, which is exactly the trend of LLM decoder in ASR tasks, the verification overhead of the target model is anticipated to become a significant bottleneck. While adaptive single-sequence prediction enhances verification efficiency by adjusting prediction lengths dynamically, further acceleration is possible, particularly in configurations characterized by a significant size disparity between the draft and target models.

\begin{figure}[!tb]
    \centerline{\includegraphics[width=0.95\linewidth]{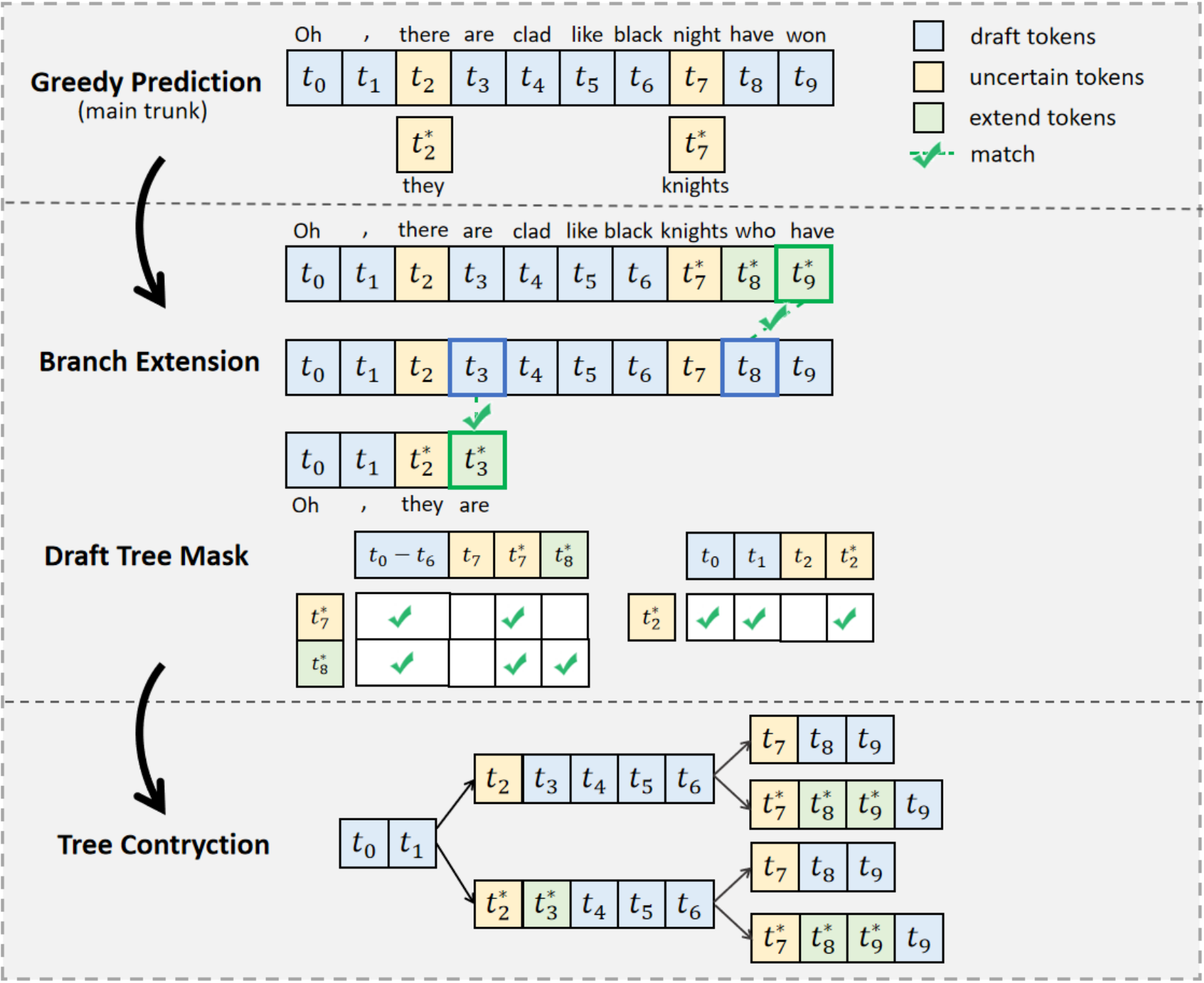}}
    \caption{Two-pass sparse-tree prediction mechanism.}
    \label{fig9}
\end{figure}

Tree-structured speculative decoding enhances verification acceptance rates by constructing a token tree within the draft model, thereby facilitating diverse predictions. However, its effectiveness is constrained by limitations on prediction length. To address this challenge, we propose a sparse token tree with limited-width expansions. As illustrated in Fig.~\ref{fig9}, the process begins with single-sequence greedy decoding, followed by the identification of potential verification failures based on output logits. Rather than truncating generation upon detecting uncertain tokens in single-sequence prediction, we mark their positions and tokens with top-k highest probability, encouraging the draft model to continue generating tokens. Upon completion of the single-sequence generation, which serves as the "main trunk" of the token tree, we perform multi-branch exploration exclusively at the points of uncertainty in the sequence. This exploration extends forward based on the candidate tokens which identified earlier, constituting the second draft decoding pass. To minimize the computational cost of this exploration, we employ the draft sequence recycling strategy. Specifically, when a branch can be concatenated to the "main trunk" or any previously generated branches, we refrain from further extending it and instead generate new branches by appending them to the "main trunk" or the existing branches. Our experiments demonstrate that extending the draft token tree by tokens with the second highest probability yields effective results, as discussed in the following section.

In the subsequent verification phase, we employ the attention mask from SpecInfer \cite{miao2023specinfer} to parallelize the verification of candidate sequences. For segments that fail verification, we apply the recycling strategy to regenerate them and mark the points of uncertainty to construct the draft token tree in the next iteration.

\section{Experimental Evaluation}
\subsection{Experimental Setup}

The proposed SpecASR is deployed on NVIDIA RTX A6000, with its performance evaluated on LibriSpeech dataset \cite{panayotov2015librispeech}. Recent studies on LLM-based ASR models often employ Llama or Vicuna \cite{chiang2023vicuna} as LLM decoders. However, ASR models based on these architectures are not yet publicly available. As an alternative, we utilize the open-source Whisper models, selecting tiny.en and medium.en versions as draft and target models, respectively, and record the decoding trajectories of Whisper-based SpecASR. Subsequently, we simulate these trajectories using TinyLlama as the draft model and Llama-7B/Vicuna-13B as target models. We observe that the word error rate (WER) gap between TinyLlama and Llama-7B/Vicuna-13B, as reported by \cite{ma2024embarrassingly}, is smaller than that between Whisper tiny.en and medium.en on the LibriSpeech dataset. This suggests that SpecASR is likely to achieve better alignment between draft and target models when deployed with Llama/Vicuna models, thereby enabling superior acceleration performance compared to our experimental results.

\subsection{Main Results}

\begin{figure}[!tb]
    \centering    
    \subfloat[][]{
	\includegraphics[width=1.\linewidth]{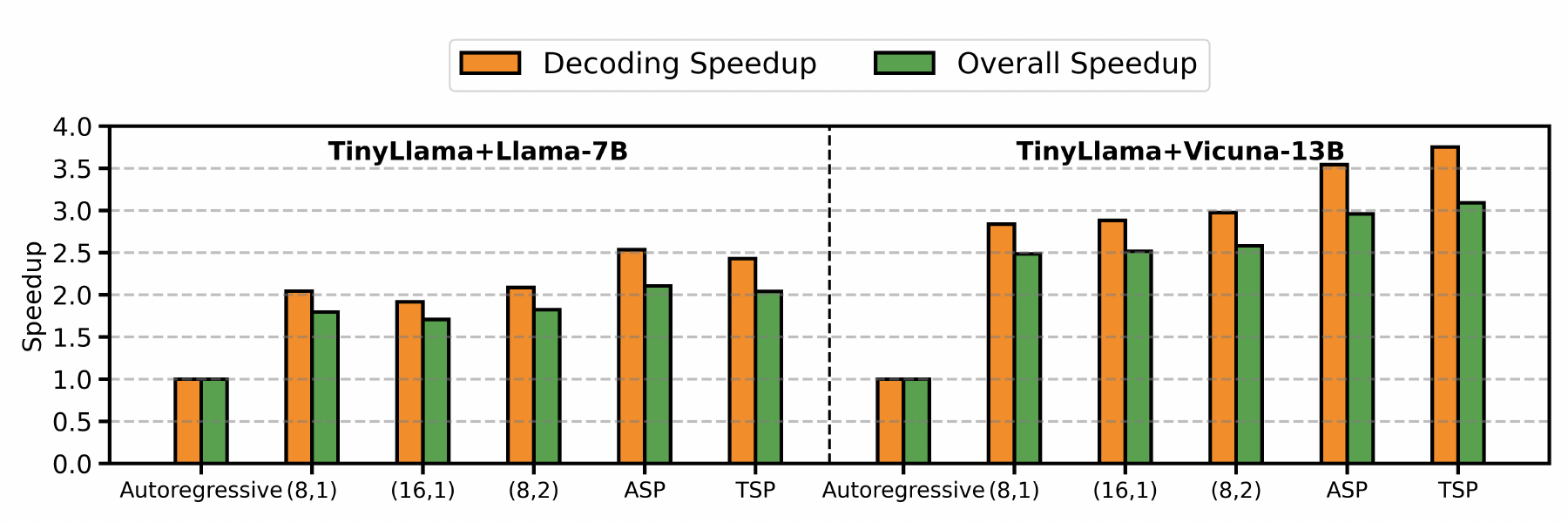}
        \label{fig:subfig10a}
    }
    \hspace{0.01\textwidth}
    \subfloat[][]{
	\includegraphics[width=1.\linewidth]{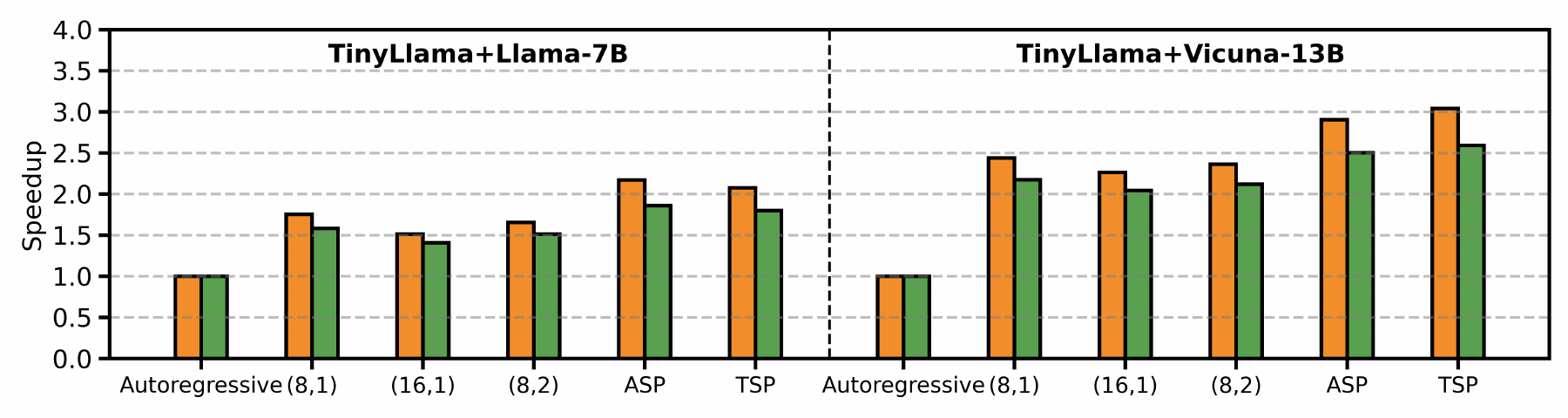}
        \label{fig:subfig10b}
    }
    \hspace{0.01\textwidth}
    \subfloat[][]{
	\includegraphics[width=1.\linewidth]{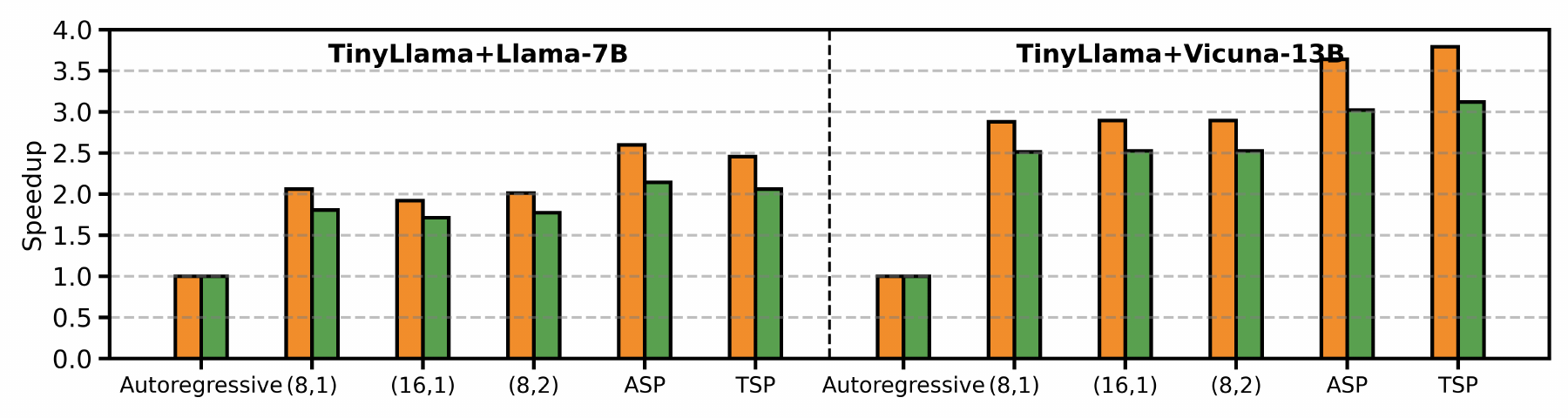}
        \label{fig:subfig10a}
    }
    \hspace{0.01\textwidth}
    \subfloat[][]{
	\includegraphics[width=1.\linewidth]{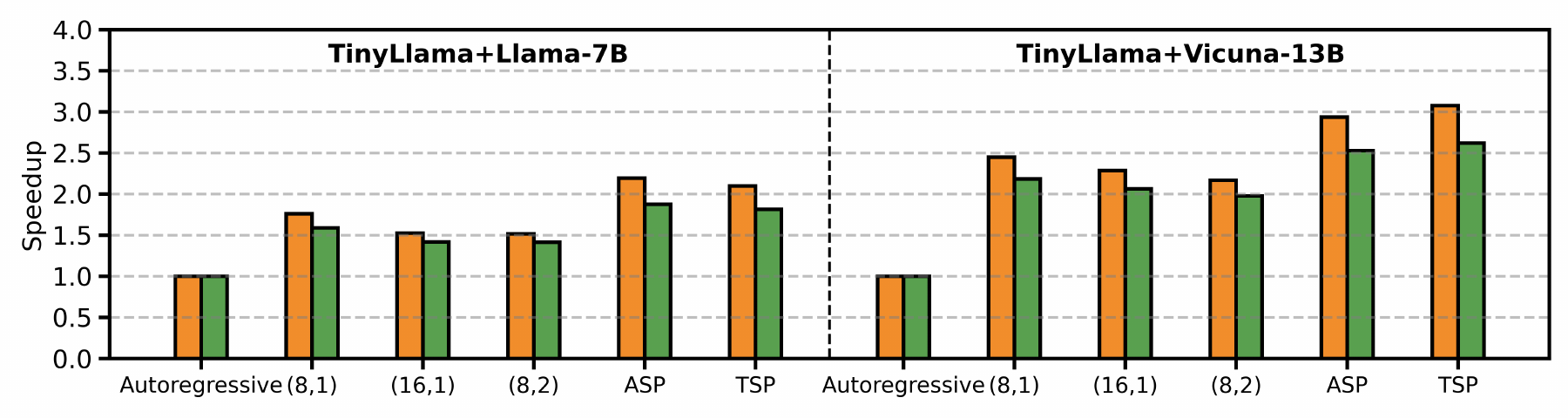}
        \label{fig:subfig10b}
    }
    \caption{Speedup comparison with baseline methods, including autoregressive decoding, speculative decoding with 8/16 prediction length and 1/2 beams (denoted as (8, 1), (16, 1), (8, 2) in the figure) on datasets: (a) test-clean, (b) test-other, (c) dev-clean, (d) dev-other.}
    \label{fig10}
\end{figure}

We evaluate SpecASR with adaptive single-sequence prediction (ASP) and two-pass sparse-tree prediction (TSP) on LibriSpeech test-clean, test-other, dev-clean, and dev-other datasets, as shown in Fig.~\ref{fig10}. For comparison, we include baseline models employing autoregressive and speculative decoding, configured with (prediction length, beam size) pairs of (8, 1), (16, 1), and (8, 2). A normalized logits threshold of 0.4 is applied to identify uncertain predicted tokens. For two-pass sparse-tree prediction, we choose to extend the draft token tree by incorporating the token with the second highest probability. Detailed discussions are provided later in this section.


\begin{figure}[!tb]
    \centering    
    \subfloat[][]{
	\includegraphics[width=0.48\linewidth]{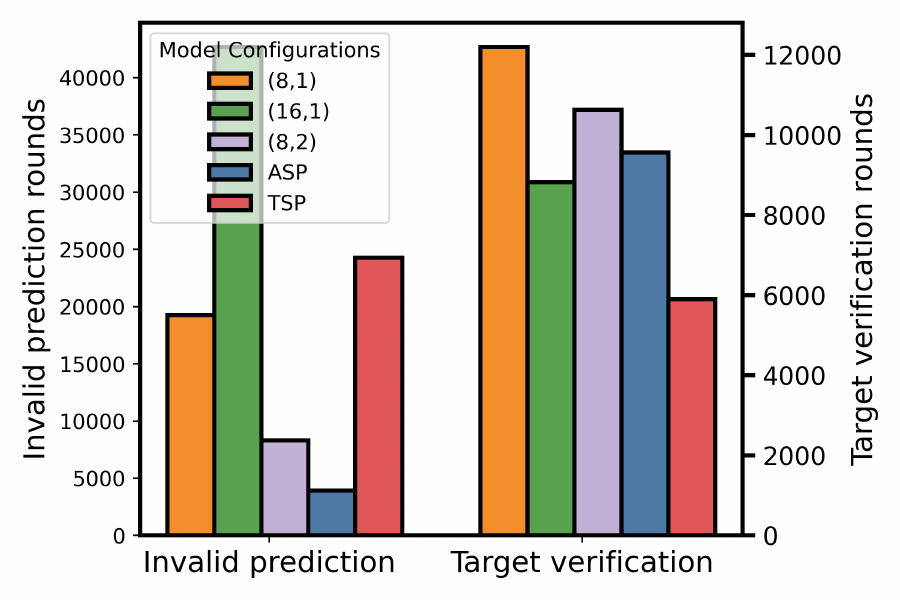}
        \label{fig:subfig11a}
    }
    \subfloat[][]{
	\includegraphics[width=0.48\linewidth]{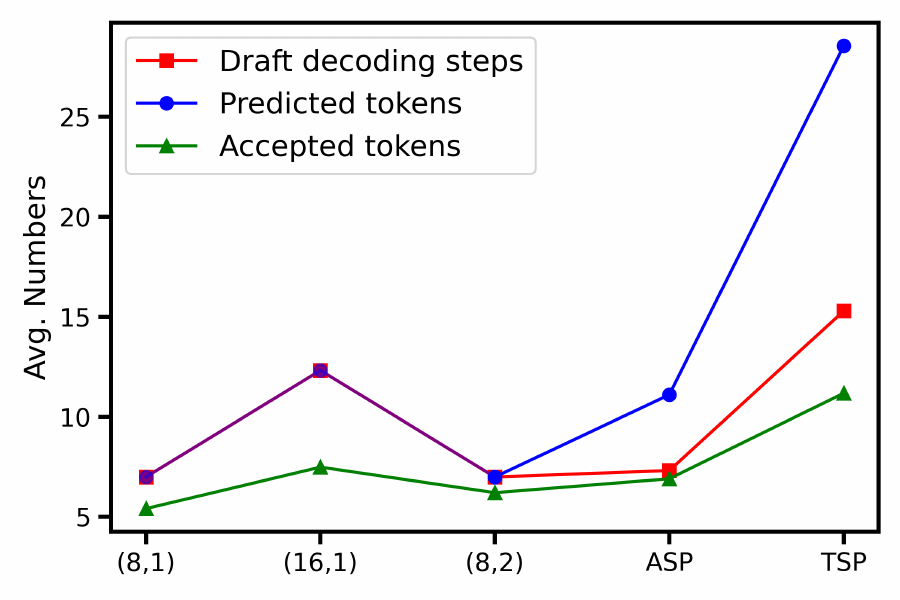}
        \label{fig:subfig11b}
    }
    \caption{Comparison of speculative methods on test-clean: (a) the number of rounds for draft prediction and target verification, (b) the average number of draft decoding steps, predicted tokens per round, and accepted tokens per round.}
    \label{fig11}
    \vspace{-10pt}
\end{figure}

\paragraph{Speedups across different model configurations and datasets} 
Using Llama-7B as the simulated target model, SpecASR achieves speedups of 2.08×–2.60× compared to autoregressive decoding and 1.21×–1.45× relative to the baseline speculative decoding. For Vicuna-13B, the speedups increase to 3.04×–3.79× over autoregressive decoding and 1.25×–1.84× over the baseline speculative decoding methods. These results underscore the versatility of the single-sequence and sparse-tree decoding strategies proposed in SpecASR, demonstrating their ability to effectively adapt to various model configurations. As illustrated in Fig.~\ref{fig11}, adaptive single-sequence prediction reduces 74.1\% of ineffective prediction steps compared to baseline speculative decoding, through early truncation and draft reuse. This results in a 94.4\% decoding-acceptance ratio per round. This approach alleviates the load on the draft model, demonstrating improved performance when the parameter disparity between the draft and target models is relatively small, as seen with TinyLlama and Llama-7B. In contrast, two-pass sparse-tree prediction enhances the diversity of the draft model's generation, leading to a slight decrease in the decoding-acceptance ratio. However, the average accepted length per verification round increases by 106.6\%, resulting in a more substantial reduction in latency, particularly when the target model is significantly larger, as in the case of Vicuna-13B. 

\paragraph{Speedup on different datasets} 
Our evaluation included noisy datasets, such as test-other and dev-other, to examine the impact of decoding accuracy on speculative efficiency. Using Vicuna-13B as the target model, SpecASR achieves a 3.04×–3.07× speedup over autoregressive decoding on these noisy datasets. Compared to the clean datasets, there is 19\% performance degradation. This is because the draft model exhibits a significant decrease in recognition accuracy compared to the target model when processing these challenging datasets. 
Consequently, the number of accepted tokens per round decreases, resulting in a lower prediction-acceptance ratio and a subsequent decline in performance. This also highlights the effectiveness of using the Whisper model to simulate Llama-based LLMs, as TinyLlama exhibits a smaller word error rate gap compared to Llama-7B and Vicuna-13B. We anticipate that SpecASR will achieve even better speedup when deployed with Llama-based ASR models.

\subsection{Impact of Key Parameters}

\begin{figure}[!tb]
    \centering    
    \subfloat[][]{
	\includegraphics[width=0.48\linewidth]{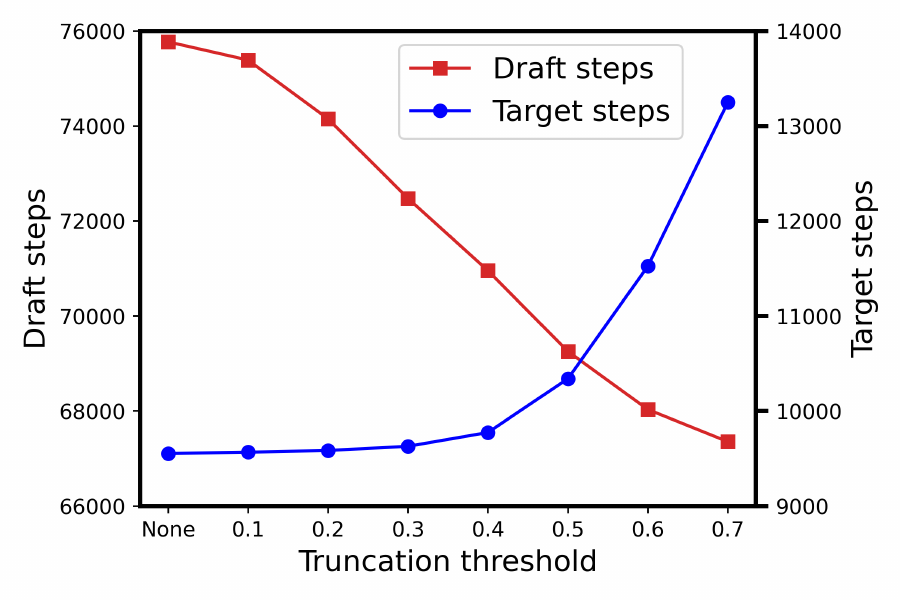}
        \label{fig:subfig12a}
    }
    \subfloat[][]{
	\includegraphics[width=0.48\linewidth]{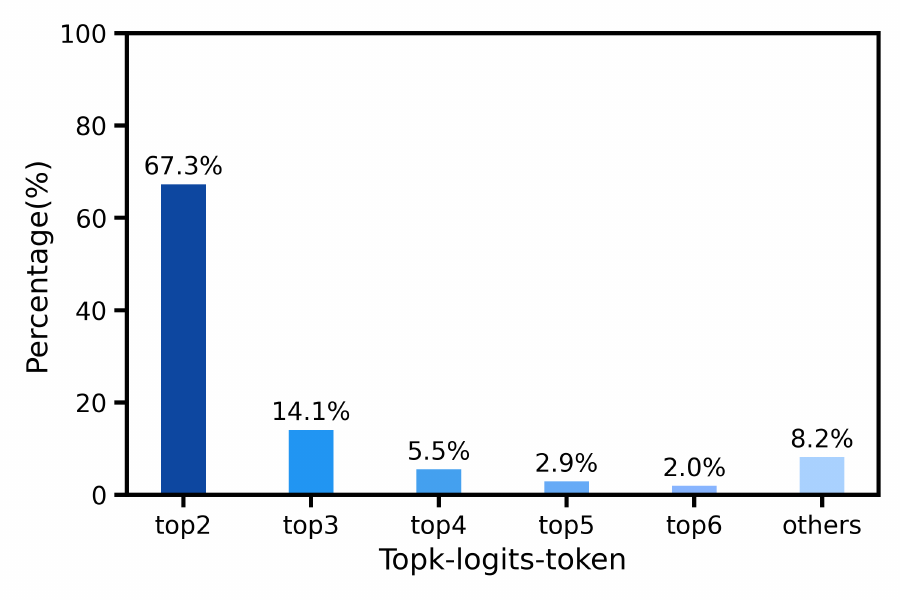}
        \label{fig:subfig12b}
    }
    \caption{(a) Step changes for draft prediction and target verification with different truncation thresholds in single-sequence speculative decoding, (b) the percentage of accepted tokens with different ranks in the draft models' output logits except for top1.}
    \label{fig12}
\end{figure}

\paragraph{Truncation threshold} 
Fig.~\ref{fig:subfig12a} illustrates the step count for both the draft and target models during single-sequence prediction with varying truncation thresholds. When the threshold is low, there are few tokens with logits below this threshold, so that the numbers of draft generation and target verification steps remains almost unchanged. As the threshold increases, the number of the draft model's generation steps decreases, while the number of the target model's verification steps increases, as some correct predictions with low logits are truncated. Further increasing the threshold leads to the truncation of additional correct predictions, causing a sharp rise in the number of target model verifications and reducing the benefits of truncation. Our experiments indicate that a threshold of 0.4 is optimal, although this may vary depending on the model.

\paragraph{Tokens with top-k highest probability} As discussed in Section IV, expanding tokens with the second highest probability yields optimal results. As shown in Fig.~\ref{fig:subfig12b}, we analyzed the rank of the token in draft model’s output that corresponds to target model's actual decoding when the top-1 token predicted fails verification. Our analysis revealed that over two-thirds of these tokens are ranked as the second-highest choice in the draft model's predictions. Given the substantial overhead of extending more branches, expanding the tree with second highest probability is the most effective solution.

\subsection{Ablation Study}
To further evaluate the effectiveness of SpecASR in accelerating ASR decoding, we progressively integrate its core techniques into the baseline speculative decoding under the Whisper tiny.en+medium.en model configurations. We then evaluate their contributions to reducing the latency of the draft model, target model, and overall system, as summarized in Tab.~\ref{tab2}. The results indicate that adaptive single-sequence prediction notably enhances the efficiency of target verification with negligible additional cost to draft model computation. By implementing the draft sequence recycling strategy, the reuse of unaccepted tokens significantly reduces draft model latency. Furthermore, due to the substantial discrepancy between the draft and target models, the two-pass sparse-tree prediction approach reduces target verification latency by over 50\% compared to the baseline speculative decoding method, while introducing only a slight increase in draft model latency, ultimately achieving the greatest speedup.

\begin{table}[!tb]
\caption{Ablation study on the average decoding latency of every 10s audio on the LibriSpeech test-clean dataset.}
\begin{center}
\setlength{\tabcolsep}{2pt}
\begin{tabular}{c|ccc}
\hline \hline
\textbf{Methods} & \textbf{Draft(ms)} & \textbf{Target(ms)} & \textbf{Total(ms)}\\
\hline
baseline speculative& 231.06 & 254.48 & 485.54\\
\hline
+adaptive single-sequence prediction& 236.23 & 191.20 & 427.43\\
\hline
+draft sequence recycling & 189.48 & 199.52 & 389.00\\
\hline
+two-pass sparse-tree prediction & 244.62 & 123.17 & 367.79\\
\hline \hline
\end{tabular}
\label{tab2}
\end{center}
\end{table}

\section{Conclusion}
In this paper, we propose SpecASR, a method designed to accelerate ASR models, particularly LLM-based ASR systems. We analyze the effectiveness of speculative decoding in ASR tasks and identify opportunities for improvement in traditional speculative decoding. Building on these observations, we introduce adaptive single-sequence prediction and two-pass sparse-tree prediction to adapt to different model configurations, maximizing speculative decoding efficiency through the recycling of draft sequences. Compared to baseline autoregressive decoding and speculative decoding, our approach achieves 3.04×–3.79× and 1.25×–1.84× speedup without compromising the recognition accuracy.


\bibliographystyle{IEEEtran}
\bibliography{reference}
\vspace{12pt}

\end{document}